\documentclass[12pt]{iopart}
\usepackage{epsf}
\newcommand{\dfrac}[2]{\displaystyle\frac{#1}{#2}}
\newcommand{\text}[1]{\mbox{#1}}

\begin{document}
\title{The Effects of Induced Scattering of Alfv\'en Waves for the Solar Wind Acceleration}
\author{A.~M.~Sadovski\dag\footnote[3]{asadovsk@classic.iki.rssi.ru}}
\date{}

\address{\dag\
Space Research Institute of Russian Academy of Sciences 117997,
Profsouznaya 84/32, Moskow, Russia}
\begin{abstract}
The analysis of energy balance of coronal holes gives that to
accelerate the fast solar wind streams the energy flux of the
order of 800~erg/cm$^2$ s is needed. Axford and McKenzie suggested
that the energy source, necessary to accelerate fast solar wind,
is in the regions of strong magnetic field that define the
boundaries of chromospheric supergranules. If the magnetic field
is not strictly unipolar in these regions, the necessary energy is
released in the processes of impulsive reconnection events. For
the plasma parameters appropriate for the coronal base such
processes of impulsive reconnection are accompanied by the
generation of Alfven and fast magnetosonic waves with the period
of the order of 1~s. On the basis of kinetic equation the model
for the evolution of Alfven waves in solar wind is suggested. The
induced scattering of these waves by plasma ions is considered as
the dominant mechanism of dissipation of Alfven waves. The
description of the evolution of the wave spectra in the process in
the processes of their propagation till the distance of 187 solar
radii was found.
\end{abstract}
\maketitle

\section{Introduction}

The solar wind which is
 formed by the gas-dynamic expansion of solar corona
 into the interplanetary space can be divided into
two states: the fast
($>700$~km/s)
and slow
($\approx 400$~km/s)
streams. The fast solar wind originates in the polar coronal
 holes,  where
 electron temperature
$T\approx 10^6$~K, plasma density $n=10^8$~cm${}^{-3}$ and
magnetic field at the coronal base
$B_0\approx10$~G~\cite{Axford},~\cite{Parker}. Variations of the
velocity of this fast solar wind streams are
 insignificant
(700--800~km/s).
The slow solar wind is associated with transient openings of
 closed field regions in the corona. The streams of slow wind
 are limited by the range
$\pm 13^{\circ}$ near the equator and the fast solar wind streams
can be found at the latitude $>\pm20^\circ$. The transition zone
between this regions
 has the width
$\pm 13^{\circ }$--$\pm20^{\circ }$ on the north and south
latitudes in the solar
 minimum~\cite{Schwenn}. The radiation balance analysis in the
 coronal holes~\cite{Withbroe} shows that to
 accelerate solar wind it is necessary to dissipate the energy
 flux of radiationless origin
$\approx8\cdot10^5$~erg/cm$^2\cdot$s within the distance of
one--two solar radii above the Sun's
 surface in order to maintain the observed temperature
$1,5\cdot 10^5$~K
of slowly rising expanding gas.

Axford and McKenzie~\cite{Axford} suggested that the energy
source, necessary to accelerate the fast solar wind streams, is in
the regions of strong magnetic field that define the boundaries of
chromospheric supergranules. If the magnetic field is not strictly
unipolar in this regions, then the necessary energy
 is released in the processes of impulsive reconnection of
 magnetic field lines on the characteristic spatial scale
 of network activity
$l=100$~km. Such processes of impulsive reconnection should be
accompanied by the
 generation of Alfv\'en and fast magnetosonic waves.
Using the parameters of plasma at the coronal base described
 above we can find the Alfv\'en wave velocity
$V_A= B/\sqrt{4\pi nm_i} = 2\cdot10^8$~cm/s,
the ion's thermal velocity
$V_{Ti}=2\cdot10^7$~cm/s,
and the characteristic time of magnetic field lines
 reconnection~\cite{Petschek}:
\begin{equation}
\tau_R =
 \frac{\pi V_A}{4l\ln \mathrm {Re_m}}
 = \frac{\pi \cdot 2 \cdot 10^8 \quad \mbox{cm/s}}
      {4 \cdot 10^7 \quad \mbox{cm}\cdot \ln 3.5 \cdot 10^{11}}
 \sim 0.6 \quad \mbox{s},
\label{1}
\end{equation}
where:
$\mathrm {Re_m} = 4 \pi n e^2 V_A l / m_e \nu_{ei} c^2
                      = 3.5 \cdot 10^{11}$
is the magnetic Reinold's number,
$\nu_{ei} = 20 n / (T^\circ)^{3/2} = 2.3$~s${}^{-1}$
is the collision frequency.
It is evident that the characteristic period of Alfv\'en waves
 is of the order of obtained reconnection time.
The total energy flux necessary to accelerate  fast solar wind
 is estimated as
$8\cdot10^5$~erg/cm${}^2\cdot$s~\cite{Axford} and the major
 part of the energy of generated waves should be dissipated
 within one-two solar radii above the solar surface.

As the dominant mechanism of dissipation of Alfv\'en waves
 generated by the processes of impulsive reconnection of
 magnetic field lines, we will consider the induced scattering
 of Alfv\'en waves by plasma ions~\cite{LeeVolk}. The great part
 of this model was developed in paper by Galeev and
 Sadovski~\cite{GalSad}
For the sake of simplicity we limit ourselves by the case of the
circularly polarized
 Alfv\'en waves propagating along the radial magnetic field
$\mathbf B_0 = B_0(r) \mathbf z$ from the polar coronal holes (the
wave vector $\mathbf k = (0,0,k)$, where $r$ is the radial
distance from the center of the Sun). For the solar wind we can
fix the following
 ordering of the ion cyclotron
$\omega_{ci}$,
Alfv\'en
$\omega_k$
and ion Doppler
$kv_{Ti}$
frequencies:
\begin{equation}
\omega_{ci}\gg\omega_k\gg kV_{Ti},
\label{2}
\end{equation}
where
$\omega_{ci}=eB_0(r)/m_ic$,
$\omega_k=kV_A$.
Under this conditions plasma is magnetized and the resonant
 condition describing the interaction of Alfv\'en waves with
 the different polarization propagating in the opposite
 directions along the magnetic field lines can be written as:
\begin{equation}
\omega_k + \omega_{k'} - \left(k+k'\right)v_{z}=0,
\label{3}
\end{equation}
where
$\omega_k =kV_A > 0$, $\omega_{k'} =-k'V_A < 0$, $k,k'>0$
and
$v_z$
is the velocity along the magnetic field.

In the next section we will describe the evolution of the Alfv\'en
waves spectrum in the processes of their propagation and show that
the wave energy flux necessary to accelerate the fast solar wind
can be dissipated at the heights of 1-2 solar radii above the
solar surface that is required by the Withbroe~\cite{Withbroe}
analysis.

\section{Evolution of Alfv\'en wave spectrum in the processes
 of their propagation}

Evolution of Alfv\'en wave amplitudes in the processes of their
 propagation into the interplanetary space we calculated
 theoretically assuming that the interplanetary magnetic field
$B_0(r)$ and the solar wind velocity $u$ are radial and
spherically symmetric, the Alfv\'en waves
 are circularly polarized, the wave amplitudes are small
$|B^\pm_k(r)|^2 k \ll B_0^2(r)$.

In this approximation we find the equation for the spectral energy
density of waves~\cite{GalSad} (see also~\cite{LivTsy},
\cite{Tsytovich}):
\begin{equation}
\left[
      u(r)+V_A(r)
\right]
\dfrac{\partial}{\partial r} \dfrac{|B^+_k(r)|^2k}{4\pi}
 = -\kappa k^2 V_A(r)
\dfrac{|B^+_k(r)|^2 k}{B_0^2(r)}
\dfrac{\partial}{\partial k} \dfrac{|B^+_k(r)|^2k}{4\pi},
\label{+1}
\end{equation}
Here unlike to previous paper~\cite{GalSad} we took bi-Maxwellian
distribution for ions and Maxwellian distribution for electrons.

Integrating equation~(\ref{+1}) we find the general
solution~\cite{Kamke} for the spectral energy density:
\begin{equation}
|B^+_k(r)|^2k = \Phi
\left[
     \kappa I \int\limits^r_{R_\odot} \dfrac{V_A(r)dr}{u+V_A(r)}
     +\dfrac{1}{k(R_\odot)} - \dfrac 1k
\right].
\label{+2}
\end{equation}
Here:
$\Phi$ is an arbitrary function;
$V_A(r) = V_A(R_\odot)R_\odot/r$;
$k(R_\odot) V_A(R_\odot)/2\pi=1.7$ Hz;
$R_\odot = 6.96 \cdot 10^{10}$ cm;
$\kappa = |B^-_k(r)|^2/|B^+_k(r)|^2$
is the ratio of the spectral energy density of Alfv\'en waves,
 propagating to the Sun and from the Sun respectively, that
 is the constant because both waves have the same spectral index
$\alpha=-1$
and radial profiles. The ratio of the wave energy density to
 the energy of the magnetic field
$I=|B^+_k(r)|^2k/B_0^2(r)$
happens to be the constant too due to the identical radial
 profiles of the energy density of Alfv\'en waves and the
 energy of magnetic field and very weak dependence of the
 wave energy density
$|B^+_k(r)|^2k$
on wave vector
$k$
due to the very small left-hand side of the equation~(\ref{+1}).
Numerical value for
$I$ can be found by  equating the energy flux of Alfv\'en
 waves generated at the base of the corona to the energy flux
 of the radiationless nature with the value
$8\cdot 10^5$~erg/cm${}^2\cdot$s,
necessary for the acceleration of the fast solar wind from
 polar coronal holes:
\begin{equation}
V_A(r)\dfrac{|B^+_k|^2k}{4\pi} = 8\cdot
10^5\text{~erg/cm}^2\cdot\text{s}. \label{+3}
\end{equation}
As the result we find that:
 $I=5\cdot 10^{-4}$.

In the process of wave propagation
 the position of the spectral break is shifted to the low
 frequencies~\cite{Tu}. The position of spectral break frequency
 is defined by the particular solution of the equation~(\ref{+2})
 which can be construct from the general solution.

We can find the exact position of the spectral break frequency
 as a function of radial distance by dividing both sides of
 this equation by the Alfv\'en wave velocity~\cite{GalSad}:
\begin{equation}
f_{br}(r) =
\left[
 0.6 +
2\pi \kappa I \int\limits^r_{R_\odot} \dfrac{V_A(r)dr}{u+V_A(r)}
\right]^{-1}\dfrac{ R_\odot }{r}.
\label{+5}
\end{equation}
Assuming that the acceleration of fast solar wind is terminated at
$r=3R_\odot$~\cite{13} we find that
$f_{br}(3R_\odot)\approx 0.28$~Hz.
At this stage most of the heating has been deposited within
 $(1-2)R_\odot$ above the surface as required by
 Withbroe~\cite{Withbroe} analysis.
We can simplify this expression at the distance
$r>10R_\odot$
assuming that the velocity of the fast solar wind
$u=7.5\cdot 10^7$~cm/s {}$\gg V_A(r)$.
As a result we reduce the expression~(\ref{+5}) to the form:
\begin{equation}
f_{br}(r) =
\left[
 0.6 +
\pi \kappa \ln\dfrac{r}{R_\odot}
\right]^{-1}\dfrac{R_\odot}{r}.
\label{+6}
\end{equation}
Taking the numerical value of the spectral break
$6\cdot 10^{-2}$ Hz
at the distance
$r=60R_\odot $ ~\cite{Tu},
separating the domain of Alfv\'en wave turbulence with the
 spectral indexes
$\alpha=-1$ and domain with $\alpha=-5/3$ and equating it to the
theoretical value obtained from
 the equation~(\ref{+6}) we find the ratio of the spectral
 energy density of Alfv\'en waves propagating to the Sun
 and from the Sun respectively that is equal to
$\kappa=0.17$.
This means that the intensity of waves propagating to the
 Sun is six times lower than those for the waves propagating
 from the Sun.

Experimental observations show that above the spectral break a
wave spectrum with a slope near $-5/3$~\cite{Bav} is
established~\cite{GalSad}:
\begin{equation}
|B^+_k(r)|^2 = |B^+_{k_{br}}(r)|^2k_{br}^{5/3}/k^{5/3}. \label{+7}
\end{equation}

The equation for the spectral energy density of waves in this case
should inherit  the structure of the equation (\ref{+1}) with the
continuos transition across the spectral break~\cite{GalSad}:
\begin{equation}
\begin{array}{l}
\left[
      u(r)+V_A(r)
\right]
\dfrac{\partial}{\partial r} \dfrac{|B^+_k(r)|^2k^{5/3}}{4\pi}\\
\qquad\qquad\quad = -\kappa k^{5/3} V_A(r)k_{br}^{-0.2}
\dfrac{|B^+_k(r)|^2 k^{5/3}}{B_0^2(r)} \dfrac{\partial}{\partial
k} \dfrac{|B^+_k(r)|^2k^{5/3}}{4\pi}, \label{+8}
\end{array}
\end{equation}
Equation for characteristics of~(\ref{+8}) take a form:
\begin{equation}
\dfrac{1}{k^{5/3}}\dfrac{d k}{d r}= \kappa I\dfrac{[2\pi
f_{br}(r)]^{2/5}V_A(r)^{3/5}}{u+V_A(r)}, \label{+9}
\end{equation}
where $I=|B^+_{k_{br}}(r)|^2k_{br}/B_0^2(r) =5\cdot 10^{-4}$,
$u\gg V_A(r)$. Beyond the $187.5R_\odot$ (0.87  astronomical units
--- AU) the collisional hydromagnetic turbulence spectrum of
Kraichnan~\cite{Krai} with the spectral index
$\alpha=-1.5$~\cite{Tu} replace the collisionless spectra with the
spectral indexes $\alpha=-1$ and $\alpha=-5/3$.

\begin{figure}
\begin{center}
\epsfbox{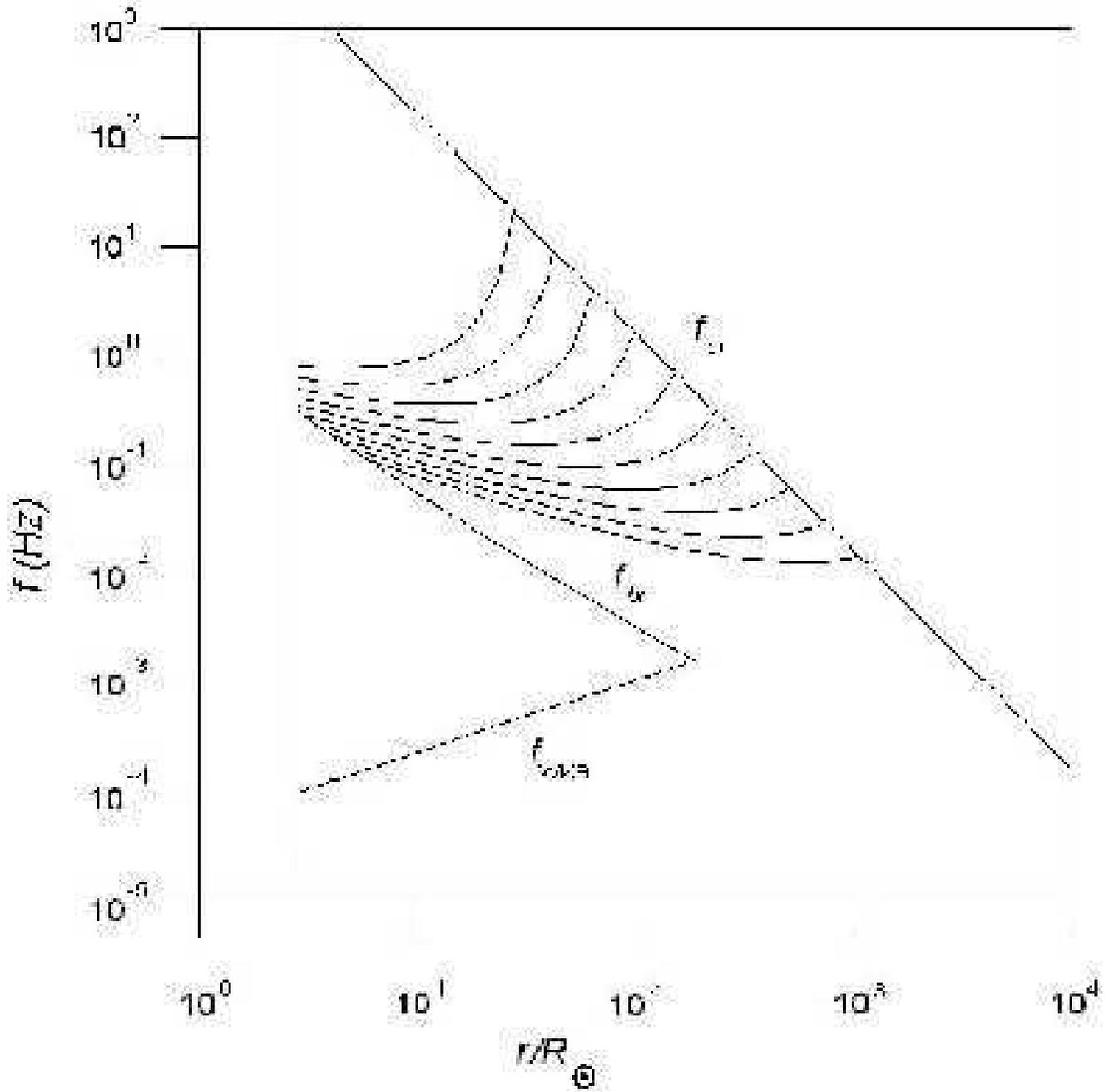}
\end{center}
\caption{\label{fig}The characteristics of the equation (30) and
the dependence of the spectral break frequency upon the distance
from the Sun}
\end{figure}

Integrating~(\ref{+9}) and transforming the result to the
 frequency dependence on the radial distance we obtain:
\begin{equation}
f=\left[
C
\left(
     \dfrac{r}{R_\odot}
\right)^{3/5}- \left(   \dfrac{1}{f_{br}(r)} \right)^{3/5}
\right]^{-5/3}, \label{+10}
\end{equation}
where $C$ is an arbitrary constant.

Figure 1 shows the characteristics of the equation~(\ref{+8})
 between the curves of the spectral break frequency
$f_{br}$
and the ion cyclotron frequency
$f_{ci}=1.6\cdot 10^4\left(R_\odot/r\right)^2$,
where the wave energy is totally absorbed in the ion cyclotron
 resonance.

To estimate the fraction of Alfv\'en wave energy transferred to
 the protons of solar wind plasma in the process of induced
 scattering of waves we use,
 following Tu~\cite{Tu1}, the equation for the magnetic moment
 of protons.
Doing this we take into account that due to the smallness of
 the left-hand side of the equation~(\ref{+8}) the product
$|B^+_k(r)|^2 k^{5/3}$ has a very weak dependence on the wave
vector $k$. Therefore the right-hand side of~(\ref{+8}) also
depends weakly on $k$ and we can rewrite it in form~\cite{GalSad}:
\begin{equation}
\left[
      u(r)+V_A(r)
\right]
\dfrac{\partial}{\partial r} \dfrac{|B^+_k(r)|^2}{4\pi}
 = -\dfrac{\partial}{\partial k} 0.25 \kappa f_{br}(r) I^2
B_0^2(r)\equiv -\dfrac{\partial}{\partial k} \Pi(r),
\label{+11}
\end{equation}
where
$\Pi(r)$
is the volumetric energy flux.

The equation for the magnetic moment take a form:
\begin{equation}
u \dfrac{d}{dr}\ln
\left(
       \dfrac{T_\perp}{B}
\right) =
\dfrac{\Pi(r)}{n(r)k_B T_\perp},
\label{+12}
\end{equation}
where
$ T_\perp $
is the perpendicular proton temperature in Kelvin,
$k_B$
 Boltzmann's constant,
$ n(r)=10^8 (R_\odot/r)^2$ cm${}^{-3}$
density of plasma.

We integrate the equation~(\ref{+12}) from the three solar
 radii assuming that the acceleration of solar wind is
 already achieved.
As a result we have:
\begin{equation}
\dfrac{T_\perp}{B_0}-\dfrac{T_\perp}{B_0}
 = \int\limits_{3R_\odot}^r
\dfrac{\Pi(r)}{n(r)k_B u B_0(r)}dr =
0.144 \ln
\left[
     0.6 +\pi\kappa\ln\dfrac{r}{R_\odot}.
\right],
\label{+13}
\end{equation}

Non-adiabatic heating of the protons related to the presence of
 the volumetric energy flux
$\Pi(r)$
reach only
$0.16$ K$/$nT
in the range of radial distances
$[3R_\odot,200R_\odot]$.
Therefore the source of energy for the acceleration of fast
 solar wind from the polar coronal holes must now be found in
 the solar corona itself~\cite{Marsch}.

Below the spectral brake frequency
$f_{br}(r)$
we have the spectrum of
 Livshits-Tsytovich~\cite{LivTsy},~\cite{Tsytovich}
 with the spectral index
$\alpha=-1$.
However this spectrum doesn't extend from
$f_{br}$
to the whole low frequency domain but truncated by the WKB
 solution.

Solving the stationary equation for Alfv\'en
waves~\cite{Whang},~\cite{Holl},~\cite{Barnes}:
\begin{equation}
\nabla
\left(\left(
      3 {\bf u}(r)\pm 2{\bf V}_A(r)
\right)
\dfrac{|B^+_k(r)|^2k}{8\pi}
\right) - {\bf u}(r)\nabla
\dfrac{|B^+_k(r)|^2 k}{8\pi}=0,
\label{+14}
\end{equation}
the equation for the wave amplitude was found in the form:
\begin{equation}
\left[
      1\pm \dfrac{V_A(r)}{u}
\right]^2 n(r)^{-3/2}|B^+_k(r)|^2k
= {\rm const}.
\label{+15}
\end{equation}
At the distances
$r>10R_\odot$
we can neglect the small parameter
$V_A(r)/u\ll 1$
and rewrite it in the form defining the reduction of the wave
 amplitudes and their spectral index
$\alpha=-2/3$~\cite{Tu}:
\begin{equation}
{\rm const}\cdot n(r) =
\left(
    |B^+_k(r)|^2k
\right)^{2/3}.
\label{+16}
\end{equation}
Besides that, knowing from observations that the spectral
 break curve
$f_{br}$
and the WKB solution are merging at the distance 0.87 AU
(187.5 $R_\odot$)
we are able to find the form and position of WKB solution
 curve from the equation~(\ref{+16}) in the form (fig.~1):
\begin{equation}
f_{WKB}(r)= 1.6\cdot 10^{-3}\cdot
\left(\dfrac{r}{187.5R_\odot}\right)^{2/3}\text{{} ľ}.
\label{+17}
\end{equation}

\section{Conclusion}

Using the  collisionless kinetic equation for
 interacting Alfv\'en waves in the random phase approximation
 we have followed analytically the evolution of these waves in the
 process of their propagation from the polar solar corona.
As a result of such evolution the boundary of waves with the
 spectral index
$\alpha=-1$
is shifted towards low frequencies due to induced scattering
 of waves by ions, which increase energy of ions by the recoil
 effect.
However, this spectrum does not extend over the whole low
 frequency range but is truncated by WKB solution.

Above the frequency $f_{br}$ the spectrum with the spectral index
$\alpha=-5/3$ is obtained. Let us note that both in the polar
solar corona and in the
 interplanetary space the approximation of collisionless
 plasma is valid at least up to the distance 0.87 AU.


\end{document}